\documentclass[11pt]{article}

\usepackage[margin=1in]{geometry}
\usepackage{amsmath,amssymb,amsthm}
\usepackage{mathtools}
\usepackage{hyperref}
\hypersetup{hidelinks}
\usepackage{cleveref}
\usepackage{xcolor}

\newtheorem{theorem}{Theorem}[section]
\newtheorem{lemma}[theorem]{Lemma}

\newtheorem{claim}[theorem]{Claim}

\newtheorem{corollary}[theorem]{Corollary}

\theoremstyle{definition}

\crefname{theorem}{theorem}{theorems}
\Crefname{theorem}{Theorem}{Theorems}
\crefname{lemma}{lemma}{lemmas}
\Crefname{lemma}{Lemma}{Lemmas}
\crefname{proposition}{proposition}{propositions}
\Crefname{proposition}{Proposition}{Propositions}
\crefname{claim}{claim}{claims}
\Crefname{claim}{Claim}{Claims}
\crefname{observation}{observation}{observations}
\Crefname{observation}{Observation}{Observations}
\crefname{corollary}{corollary}{corollaries}
\Crefname{corollary}{Corollary}{Corollaries}
\crefname{definition}{definition}{definitions}
\Crefname{definition}{Definition}{Definitions}
\crefname{remark}{remark}{remarks}
\Crefname{remark}{Remark}{Remarks}

\newcommand{\R}{\mathbb{R}}

\newcommand{\E}{\mathbb{E}}

\newcommand{\1}{\mathbf{1}}

\newcommand{\cA}{{\mathcal A}}
\newcommand{\cC}{{\mathcal C}}
\newcommand{\cD}{{\mathcal D}}
\newcommand{\cU}{{\mathcal U}}

\newcommand{\cQ}{{\mathcal Q}}

\newcommand{\cE}{{\mathcal E}}
\newcommand{\cF}{{\mathcal F}}
\newcommand{\ba}{{\mathbf a}}
\newcommand{\be}{{\mathbf e}}
\newcommand{\bx}{{\mathbf x}}

\newcommand{\bv}{{\mathbf v}}
\newcommand{\bV}{{\mathbf V}}

\newcommand{\bG}{{\mathbf G}}

\newcommand{\by}{{\mathbf y}}
\newcommand{\bu}{{\mathbf u}}

\newcommand{\bW}{{\mathbf W}}

\newcommand{\bz}{{\mathbf z}}
\newcommand{\bU}{{\mathbf U}}
\newcommand{\bX}{{\mathbf X}}
\newcommand{\bY}{{\mathbf Y}}
\newcommand{\bZ}{{\mathbf Z}}
\newcommand{\bone}{{\mathbf 1}}
\newcommand{\bSigma}{{\mathbf \Sigma}}
\newcommand{\OPT}{\mathrm{OPT}_f}

\renewcommand{\epsilon}{\varepsilon}
\renewcommand{\phi}{\varphi}

\newcommand{\IE}{\mathbb{E}}

\newcommand{\IR}{\mathbb{R}}

\newcommand{\IP}{\mathbb{P}}
\newcommand{\IZ}{\mathbb{Z}}

\DeclareMathOperator{\poly}{poly}

\DeclareMathOperator{\supp}{supp}
\DeclareMathOperator{\Var}{Var}
\DeclareMathOperator{\Cov}{Cov}

\newcommand{\ru}[1]{}
\newcommand{\jv}[1]{}
\newcommand{\mar}[1]{}

\title{Optimal universal sets for submodular functions}
\author{Marek Adamczyk \\ Koźmiński University \and
 Richard Ueltzen \\ Stanford University \and
Jan Vondr\'ak \\ Stanford University}

\begin{document}

\maketitle

\begin{abstract}
We resolve the question of deterministic non-adaptive algorithms for nonnegative submodular maximization: We prove that for every $n \geq 2$ and $\epsilon \in (0,\frac{1}{4})$, there is a polynomial-size family $\cF_{n,\epsilon}$, $|\cF_{n,\epsilon}| = n^{O(1/\epsilon^2 \log (1/\epsilon))}$, such that for every nonnegative submodular function $f:2^{[n]} \to \R$, $\max_{S \in \cF_{n,\epsilon}} f(S) \geq (1/4 - \epsilon) \max_{S \subseteq [n]} f(S)$. It is also easy to show that a factor better than $1/4$ cannot be achieved by any fixed polynomial-size family.

A related result that we prove is that there is a distribution $\cD_{n,\epsilon}$ supported on polynomially many subsets of $[n]$ such that $(1 - \epsilon) \E_{R \sim \cU_n}[f(R)] \leq \E_{R \sim \cD_{n,\epsilon}}[f(R)] \leq \E_{R \sim \cU_n}[f(R)]$, where $\cU_n$ is the uniform distribution on $2^{[n]}$, and $f$ is any nonnegative submodular function. This is interesting in light of the fact that uniform convergence does not hold for submodular functions: Sampling polynomially many sets from $\cU_n$ and taking the empirical mean does not converge to the true expectation uniformly for all submodular functions.

The construction of $\cF_{n,\epsilon}$ and $\cD_{n,\epsilon}$ is based on a new semidefinite extension of submodular functions: We show that the Goemans-Williamson rounding technique for a near-orthogonal arrangement of vectors gives a value close to the uniform expectation, and this rounding can be derandomized, since near-orthogonal vectors can be constructed in logarithmic dimension.
\end{abstract}

\section{Introduction}

Submodular functions\footnote{$f:2^V \to \R$ is submodular if $f(S \cup T) + f(S \cap T) \leq f(S) + f(T)$ for all $S,T \subseteq V$.} capture a discrete notion of convexity which has led to abstract formulations and generalizations of classical combinatorial optimization problems, involving cuts, coverage, spanning trees, matroids, etc. In this paper, we consider arguably the most basic submodular maximization problem: unconstrained maximization of a nonnegative submodular function $f(S)$ over $S \subseteq [n]$. This generalizes the well-known \textsc{Max-Cut} problem, where $f(S)$ is the cut function in a given graph.

The work of Feige et al.~\cite{FeigeMV07} gave the first non-trivial results  for unconstrained submodular maximization: a deterministic $1/3$-approximation algorithm and a randomized $2/5$-approximation algorithm for every nonnegative submodular function. They also proved that any $(1/2+\epsilon)$-approximation requires an exponential number of queries in the value oracle model.
Buchbinder et al. \cite{BuchbinderFNS15} later closed the gap and designed a randomized algorithm achieving the optimal approximation ratio of $1/2$. Also, Buchbinder and Feldman \cite{BuchbinderFeldman15} showed that the same approximation ratio of $1/2$ can be achieved deterministically.

However, the approximation ratio is only one part of the picture. The algorithms above are \emph{adaptive}: they query $f$ on partial solutions and use the answers to guide their next choices. The same paper of Feige et al.~\cite{FeigeMV07} also gives a guarantee that does away with such queries: if $R$ includes each element independently with probability $1/2$, then
\[
\mathbb{E}[f(R)] \geq \frac{1}{4}\max_{S\subseteq[n]} f(S).
\]
The factor is weaker, but this algorithm has a very special feature: it never looks at $f$. It needs only the ground-set size $n$ and $n$ independent random bits, and yet it gives an approximation guarantee of $\frac{1}{4}$  for every nonnegative submodular function. We call this an \emph{oblivious} algorithm: it chooses the solution without learning anything about the function it will be evaluated on.

More generally, there has been interest in designing algorithms which access the objective function in a limited fashion, by issuing only a certain number of rounds of queries. 
This line of research, known as {\em adaptive complexity} of optimization problems, was initiated by Balkanski and Singer \cite{BalkanskiS18}. In this model, we access an objective function $f(S)$ through a sequence of parallel query rounds $Q_1,Q_2,\ldots$, where all queries in each round are executed in parallel and may depend only on the answers to previous rounds. Ideally, the number of rounds is small, constant or logarithmic in the input size. Adaptive complexity is the number of rounds needed to achieve a specified approximation guarantee.
The motivation comes from applications where access to $f(S)$ is the bottleneck, but we assume that many evaluations of $f(S)$ can be carried out in parallel. The model was first studied for monotone submodular maximization under a cardinality constraint $|S| \leq k$, where a $(1-1/e-\epsilon)$-approximation can be achieved in $O(\log n\log k\,\mathrm{poly}(1/\epsilon))$ rounds \cite{BalkanskiRS19a}. A nearly logarithmic number of rounds is necessary for a cardinality constraint \cite{BalkanskiS18}, and subsequent work extended this study to other constraints \cite{balkanski2019optimal,chekuri2019parallelizing,ene2019submodular,li2020polynomial}.
For our problem of interest, unconstrained submodular maximization, it was shown that $O(1/\epsilon)$ rounds of queries are sufficient to achieve a $(1/2 - \epsilon)$-approximation \cite{ChenFK19}. 

This raises a natural question: Can a constant factor be achieved in 1 round of queries, without any adaptivity at all? With randomization, this is possible as we mentioned above, but deterministically it is unclear and this question has been open for at least 15 years (since it was explicitly raised in \cite{FeigeMV11}). More specifically, we ask the following.

\medskip

\emph{Is there a deterministic non-adaptive algorithm that, given only the input size $n$, queries $f(S)$ on a fixed universal family $\cF_n\subset 2^{[n]}$ of polynomially many sets, and guarantees $\max_{S\in\cF_n} f(S) \geq c \cdot \max_{T\subseteq[n]} f(T)$ for every nonnegative submodular function $f$ and a universal constant $c>0$?}

\medskip

This question is especially intriguing in light of Luby's 1988 result for \textsc{Max-Cut}: not only can the problem be approximated obliviously, it can be done deterministically, with no random bits at all \cite{Luby88,VadhanPseudorandomness}. Given only $n=|V|$, one can fix a family of $2(n+1)$ subsets $S_1,\ldots,S_{2(n+1)}\subseteq V$ before seeing the graph, so that for every graph $G$ on $V$,
\[
\max_{i\in[2(n+1)]}\mathrm{CUT}_G(S_i)\geq \frac{1}{2}\max_{S\subseteq V}\mathrm{CUT}_G(S).
\]
%\ru{Maybe drop "pseudorandom" in the following sentence? I guess pairwise independence implies some pseudorandomness, but the formulation sounds like we need pairwise indpeendence AND pseudorandomness.}
Moreover, the $2(n+1)$-sized family of Luby is actually a support of a pairwise independent distribution over subsets of $V$. Thus, evaluating the cut function on a fixed family of $2(n+1)$ sets gives a deterministic non-adaptive $1/2$-approximation for every (weighted) graph on $n$ vertices.
%\jv{I prefer to say "non-adaptive" as opposed to "oblivious". By oblivious I would understand an algorithm like the randomized $1/4$, which doesn't look at any values at all. Here, we check $O(n)$ values and return the maximum.}

Whether this generalizes to submodular functions was an open question. 
Recent work \cite{ChekuriUV26} gave the first non-trivial results for this problem. There is a universal family $\cF\subseteq 2^{[n]}$ of $2\lceil\log_2 n\rceil$ subsets such that, for every nonnegative submodular function $f$, the maximum over this family achieves at least a $1/(2\lceil\log_2 n\rceil)$ fraction of $\max_{S\subseteq[n]} f(S)$. Further, there is a family of size $n^{o(1)}$ that gives an $\Omega(\log\log n/\log n)$-approximation. 
%The same work also answered this question for ``absolute non-monotone XOS functions'', for which the optimal factor is $\Theta(\sqrt{\log n/n})$. 
But this paper also showed that pairwise independence, or even $k$-wise independence for constant $k$, is not sufficient for submodular functions, and sometimes gives only an $O({1}/{\sqrt{\log n}})$-approximation.

\subsection{Our results}

In this paper, we resolve the question of deterministic non-adaptive maximization of submodular functions.

\begin{theorem}
\label{thm:main}
For every $\epsilon \in (0, \frac{1}{4})$ and every $n \geq 2$, there is a family of sets $\cF_{n,\epsilon} \subset 2^{[n]}$, $|\cF_{n,\epsilon}| = n^{O(1/\epsilon^2 \log (1/\epsilon))}$, such that for every nonnegative submodular function $f:2^{[n]} \to \R$,
$$ \max_{S \in \cF_{n,\epsilon}} f(S) \geq \left( \frac{1}{4} - \epsilon \right) \max_{S \subseteq [n]} f(S). $$
\end{theorem}

{\em Remark:} It was already known \cite{FeigeMV07} that a factor better than $1/4$ cannot be achieved by a universal family. More precisely, for any $\epsilon>0$, if $\cF \subset 2^{[n]}$ is such that for every nonnegative submodular function $f:2^{[n]} \to \R$,
$ \max_{S \in \cF} f(S) \geq \left( \frac{1}{4} + \epsilon \right) \max_{S \subseteq [n]} f(S),$
then $\cF$ must be exponentially large in $n$.

%It is relatively easy to prove that a factor better than $1/4$ cannot be achieved by any polynomial-size family.
%\ru{I find the part in parantheses above not super coherent, and I would like it to be edited or removed. The strong concentration only happens for the particular construction (any balanced directed cut) showing that $1 / 4$ is optimal for a uniformly random set.}
%\jv{OK, I removed it.}

%\begin{theorem}
%\label{thm:converse-to-main}
%    For every $\epsilon \in (0, 1 / 2)$ and every $n \geq 1$, and any family of sets $\cF \subset 2^{[n]}$ such that for every nonnegative submodular function $f:2^{[n]} \to \R$,
%    $$ \max_{S \in \cF} f(S) \geq \left( \frac{1}{4} + \epsilon \right) \max_{S \subseteq [n]} f(S), $$
%    it holds that $|\cF| \geq e^{\epsilon^2 n / 2} - 2$.
%\end{theorem}

%For completeness, we present a short proof in \Cref{proof-of-converse-to-main-thm}.
%\ru{It is still open to figure out a polynomial lower bound on $|\cF|$ for a $(1 / 4 - \epsilon)$-approximation.}
%\jv{We could mention it in the conclusions; here I don't think we need to discuss the remaining gaps.}
%\ru{Ok.}

We also provide a closely related result for the estimation of the uniform expectation of a submodular function by a distribution of polynomial-size support. 

\begin{theorem}
\label{thm:main-exp}
For every $\epsilon \in (0, 1)$ and $n \geq 2$, there is a distribution $\cD_{n,\epsilon}$ over subsets of $[n]$, with support of size $n^{O(1/\epsilon^4 \log (1/\epsilon))}$, such that for every nonnegative submodular function $f:2^{[n]} \to \R$,
$$ \E_{R \sim \cU_n}[f(R)] \geq \E_{R \sim \cD_{n,\epsilon}}[f(R)] \geq \left( 1 - \epsilon \right) \E_{R \sim \cU_n}[f(R)] $$
where $\cU_n$ is the uniform distribution over subsets of $[n]$.
\end{theorem}

Interestingly, the family $\cF_{n, \epsilon}$ and the related distribution $\cD_{n,\epsilon}$ are {\em not} obtained by sampling independent random subsets of $[n]$. In fact, we still do not know whether a sample of polynomially many independent random sets achieves the bound of Theorem~\ref{thm:main}. Moreover, we do know that by sampling independent random sets we {\em cannot} achieve the result of Theorem~\ref{thm:main-exp}.

\begin{theorem}
\label{thm:no-uniform-convergence}
Let $\delta>0$ and let $\cF_n$ be a family of polynomially many independently and uniformly random subsets of $[n]$. Then with high probability, there is a nonnegative submodular function $f:2^{[n]} \to \R_{\geq 0}$ such that
$$ \max_{S \in \cF_n} f(S) < \left( \frac{1}{2} + \delta \right) \E_{R \sim \cU_n}[f(R)].$$
\end{theorem}

This can be interpreted in the context of {\em uniform convergence} in machine learning:
One can ask whether taking an empirical average over a family of independently sampled subsets from $\cU_n$ is a good estimate of the true expectation over $\cU_n$, simultaneously for all nonnegative submodular functions $f$. As Theorem~\ref{thm:no-uniform-convergence} shows, this is not true: uniform convergence {\em does not hold} for submodular functions under the uniform distribution. However, interestingly, we can obtain a good estimate of $\E_{R \sim \cU_n}[f(R)]$ by sampling from a different distribution $\cD_{n,\epsilon}$, provided by Theorem~\ref{thm:main-exp}.
%The distribution $\cD_{n,\epsilon}$ has polynomial-size support and hence sampling from it polynomially many times gives a good estimate of the expectation over $\cD_{n,\epsilon}$ (and hence also over $\cU_n$), simultaneously for all nonnegative submodular functions. 
We discuss this in more detail in Section~\ref{sec:uniform-convergence}.

\subsection{AI usage statement}

The construction of universal families proving Theorem~\ref{thm:main} was found using extensive interactions with ChatGPT-6 Astra. The exposition of this result in the context of a semidefinite extension of submodular functions, as well as Theorem~\ref{thm:main-exp} and its interpretation in the context of uniform convergence, are due to the authors. The negative result of Theorem~\ref{thm:no-uniform-convergence} was known to the authors earlier, with a weaker factor of $3/4$. The factor in Theorem~\ref{thm:no-uniform-convergence} was improved from $3/4$ to $1/2$ by ChatGPT-6 Astra. 

\subsection{Organization of the paper}

In Section~\ref{sec:outline}, we provide an outline of our approach without technical proofs. First, Section~\ref{sec:overview} gives a high-level explanation, and then Sections~\ref{sec:semidefinite-extension} and \ref{sec:derandomization} provide some more details. In Section~\ref{sec:uniform-convergence}, we discuss the issue of uniform convergence and why our approach allows one to estimate the expectation uniformly, while sampling from the uniform distribution doesn't. In Section~\ref{sec:proofs}, we provide the formal proofs, with some proofs of known or folklore results deferred to Appendix~\ref{app:proofs}.

\section{Outline of the Approach}
\label{sec:outline}

As we discussed above, universal families exist for the \textsc{Max-Cut} problem due to the fact that pairwise independence is sufficient to recover $1/2$ of the optimal value. Unfortunately, this does not extend to submodular functions and even $k$-wise independence (for any fixed $k$) is not sufficient to achieve a constant factor times the optimum \cite{ChekuriUV26}.
Still, the conceptual connection with the \textsc{Max-Cut} problem turns out to be useful. Recall the celebrated Goemans-Williamson algorithm for \textsc{Max-Cut} \cite{GoemansW95}: First, they consider a semidefinite relaxation where each element is represented by a vector $\bv_i$, and a solution is produced by separating the vectors into two disjoint sets by a random hyperplane. It is an intriguing question whether this relaxation can be lifted somehow to submodular functions. While we do not have a completely satisfying answer to this question in general, it turns out that an approach based on a semidefinite formulation leads to a solution of our problem.

\subsection{Overview of the Technique}
\label{sec:overview}

We begin with a classical bound for independent sampling,
which will be used repeatedly throughout our construction.
\begin{lemma}\label{submodular-on-p-biased-set-approximates-maximum}
    If $R_p\subseteq[n]$ contains each element independently
    with probability $p\in[0,1]$, then for every nonnegative
    submodular function $f : 2^{[n]} \to \IR_{\geq 0}$,
    \[
    \mathbb E[f(R_p)] \geq p(1-p) \max_{S \subseteq [n]} f(S).
    \]
\end{lemma}
In particular, taking $p=1/2$ yields the familiar $1/4$-approximation of Feige et al.~\cite{FeigeMV07}.
In the following, we denote $\OPT = \max_{S \subseteq [n]} f(S)$.

\paragraph{Random hyperplane rounding.}
Our starting point is a geometric interpretation of this independent sampling guarantee. Consider the standard orthonormal basis $\be_1,\ldots,\be_n$ of $\mathbb R^n$ (or in fact any other orthonormal basis). Picking a random hyperplane with normal $\bZ \sim N(0,I_n)$ and selecting the vectors on its positive side gives a random set $R_\bZ=\{i: \langle \be_i,\bZ \rangle \geq 0\}$. In this case, $G_i = \langle \be_i, \bZ \rangle$ are independent standard Gaussians, and so $R_\bZ$ is a uniformly random subset of $[n]$ and $\mathbb \E_\bZ[f(R_\bZ)]\geq\tfrac14\mathrm{OPT}_f$ by \Cref{submodular-on-p-biased-set-approximates-maximum} with $p=1/2$.
If we could somehow emulate the distribution of this random hyperplane by polynomially many hyperplanes, then we would be done. It's not clear how to do this in $\R^n$; however, the situation is easier if the vectors are embedded in a lower dimension.

\paragraph{Low-dimension embedding.}
We replace the orthogonal basis by nearly orthogonal unit vectors $\bv_1,\ldots,\bv_n\in\mathbb \R^k$, where $k=O(\varepsilon^{-2}\log n)$, satisfying
\[
|\langle \bv_i,\bv_j\rangle| \leq \varepsilon \qquad(\forall i\neq j).
\]
This can be viewed as a Johnson-Lindenstrauss projection of an orthonormal basis, or a direct construction in $\R^k$; we show this simple result in \Cref{lem:near-orthogonal}.
In $\R^k$, we apply random-hyperplane rounding to the nearly orthogonal vectors $\bv_i$: We sample a normal vector $\bZ \sim N(0,I_k) \in \mathbb{R}^k$, and select $R_\bZ=\{i:\langle \bv_i,\bZ \rangle \geq 0\}$.
Unlike the case of an orthonormal basis,
the projections $G_i:=\langle \bv_i, \bZ \rangle$ are no longer independent. 
Instead, they form a jointly Gaussian vector with covariance
$\operatorname{Cov}[G_i,G_j]=\langle \bv_i,\bv_j\rangle$. Since the random set
$R_\bZ=\{i:\langle \bv_i,\bZ \rangle \geq 0\}$ exhibits correlations between elements, it seems non-trivial to estimate its expected value under a submodular function $f$ (we cannot use \Cref{submodular-on-p-biased-set-approximates-maximum} directly). We note that the inner products $\langle \bv_i, \bv_j \rangle$ appear to give information about {\em pairwise correlations} between elements, which we already know to be insufficient for achieving a constant factor, even with pairwise independence \cite{ChekuriUV26}.
%\ru{"... since even pairwise independence is not sufficient for approximating the maximum of a submodular function to a constant factor."}
However, it turns out that the Gaussian structure of $\bG = (G_1,\ldots,G_n)$ is more powerful and allows us to make the desired comparison.

\iffalse
\paragraph{Auxiliary gaussians.}
To remedy this, we introduce an idealized Gaussian vector $Y$ with covariance $\operatorname{Cov}(Y_i,Y_j)=\varepsilon$ for $i\neq j$:
\[
Y_i=\sqrt{1-\varepsilon}\,Z_i+\sqrt{\varepsilon}\,Z_0,
\]
where the $Z_i$ are independent standard Gaussians. The $Y$ vector has the same marginal distributions as $X$. It still has correlated coordinates, but its structure is simpler than that of $X$. In particular, the $Y_i$ are independent conditioned on $Z_0$, allowing us to apply
\Cref{submodular-on-p-biased-set-approximates-maximum}
conditionally, obtaining
\[
\mathbb E[f(R_Y) ]
\geq \mathbb E[p(Z_0)(1-p(Z_0))]\mathrm{OPT}_f,
\]
where $p(t)=\Phi(\sqrt{\varepsilon/(1-\varepsilon)} \cdot t)$
and $\Phi$ denotes the standard Gaussian CDF.
Calculating the expected value over $Z_0$ gives
$\mathbb E[p(Z_0)(1-p(Z_0))]
=1/4-\mathbb E[(p(Z_0)-1/2)^2]
=1/4-O(\varepsilon)$,
giving the desired approximation guarantee.
\fi

\paragraph{Stochastic comparison for submodular functions.}
A key inequality that can be derived from several papers in the probability literature 
\cite{HoudrePS98,Muller01,Voigtlander21} is the following: \\
{\em If $\bG$ and $\bG'$ are jointly Gaussian vectors with the same marginals $G_i \stackrel{\mathrm d}{=} G'_i$,
and $\operatorname{Cov}(G_i,G_j) \leq \operatorname{Cov}(G'_i,G'_j)$ for all $i\neq j$,
then for every nonnegative submodular function $f$, we have
$$\E[f(\{i:G_i\geq0\})] \geq \E[f(\{i:G'_i\geq0\})].$$}
In our setting, we can apply this to the Gaussian variables $G_i = \langle \bv_i, \bZ \rangle$ as above, and  $G'_i = \sqrt{1-\varepsilon}\,Y_i + \sqrt{\varepsilon} \,Y_0$,
where $Y_0,Y_1,\ldots,Y_n$ are independent standard Gaussians. 
Note that $G'_i \sim N(0,1)$ and $\operatorname{Cov}(G'_i,G'_j) = \varepsilon \geq \langle \bv_i,\bv_j\rangle = \operatorname{Cov}(G_i,G_j)$;
hence $\E[f(\{i:G_i\geq0\})] \geq \E[f(\{i: G'_i \geq 0\})]$ by the stochastic comparison inequality.

The variables $G'_i$ conditioned on $Y_0$ are independently distributed, and hence $\{i : G_i' \geq 0\}$ conditioned on $Y_0$ has a product distribution, to which we can apply \Cref{submodular-on-p-biased-set-approximates-maximum}. It follows that $\E[f(R_\bZ)] \geq \E[p(Y_0) ( 1 -p(Y_0))] \OPT \geq (1/4 - \varepsilon) \OPT$, where $p(t) = \Phi\left(\sqrt{\frac{\varepsilon}{1-\varepsilon}} \, t\right)$ and $\Phi(t) = \int_{-\infty}^t \frac{1}{\sqrt{2\pi}} e^{-s^2 / 2} ds$ is the CDF of the standard Gaussian.
In the end, the estimate we need to prove entails some additional technicalities, but it is based on the idea presented here.

\paragraph{Discretization.}
To obtain a polynomial-size universal family, we need to discretize
the space of possible solutions obtained by Gaussian rounding in $\R^k$.
\iffalse
We introduce an independent Gaussian threshold
$T\sim N(0,\varepsilon)$ and consider the shifted rounding
$R_g(T)=\{i:\langle v_i,g\rangle\geq T\}$,
where $v_1,\ldots,v_n\in\mathbb R^k$ and
$g\sim N(0,I_k)$.
\par
Crucially, our Gaussian comparison argument continues
to work after conditioning on $T$. Indeed, conditioning
further on $Z_0$, the auxiliary inclusion decisions remain
independent, now with probability
$p(T,Z_0)=\Phi((\sqrt{\varepsilon}Z_0-T)/\sqrt{1-\varepsilon})$.
Consequently,
\[
\begin{aligned}
\mathbb E_{g,T}[f(R_g(T))]
&\geq \mathbb E_{Y,T}[f(\{i:Y_i\geq T\})]\\
&\geq \mathbb E_{T,Z_0}[p(T,Z_0)(1-p(T,Z_0))]
\mathrm{OPT}_f\\
&\geq \left(\frac14-O(\varepsilon)\right)\mathrm{OPT}_f.
\end{aligned}
\]
The last inequality follows from the same Gaussian CDF
estimate as before, since
$\mathbb E[(T-\sqrt{\varepsilon}Z_0)^2]=2\varepsilon$.
This shifted-rounding guarantee is formalized
in \Cref{thm:gaussian-rounding}.
\par
\fi
We discretize the space of hyperplane normals $\bZ$ using a fine grid in
$\mathbb R^k$, and also introduce a random rounding threshold $T$
in order to be able to bound the discretization error.
We round each coordinate of $\bZ$
randomly to one of its two neighboring grid points, with
probabilities chosen so that its expected value remains preserved.
The random threshold $T$ smooths the selection rule,
allowing submodularity to control the resulting error
independently of the dimension $k$. % as quantified in \Cref{lem:discretization}.
After dropping unusually large outcomes of the discretized normal vector, we can enumerate the discretized hyperplanes and their
threshold sets to obtain a universal family of size
$n^{O(\varepsilon^{-2}\log(1/\varepsilon))}$,
achieving the desired $(1/4-\varepsilon)$-approximation,
as stated in \Cref{thm:main}.

Next, we describe these steps in more detail, and state rigorous lemmas
which we finally prove in Section~\ref{sec:proofs}.

\subsection{Semidefinite extension and Gaussian rounding}
\label{sec:semidefinite-extension}

Let us now describe more explicitly the semidefinite formulation of our problem. Given a ground set $[n]$, we represent each element $i \in [n]$ by a unit vector $\bv_i \in \R^n$, and we denote the associated Gram matrix by $\bSigma$, with entries $\sigma_{ij} = \bv_i \cdot \bv_j$.
We have $\bSigma \in \cE_n$, where $\cE_n$ denotes the collection of all $n \times n$ positive semidefinite matrices with diagonal entries all $1$.
Given a submodular function $f:2^{[n]} \to \R$, we would like to define $F:\cE_n \to \R$ which is in some sense an extension of the submodular function $f$ analogous to the SDP relaxation for \textsc{Max-Cut}. Unfortunately, it is not clear how to define such a function $F$ in a way that allows us to maximize $F$ algorithmically. However, it turns out that this is not required for our result here.

%\jv{The extension including shifted thresholds does not appear in the relaxation for Max-Cut. I think it's easier to motivate $F(\bSigma)$ first, and then define $F(\bx,\bSigma)$ as another generalization.}
%\ru{Ok. Sounds good.}
Instead, we define $F:\cE_n \to \R$ directly as the expectation of a certain randomized rounding procedure.
The rounding procedure is exactly the Goemans-Williamson random-hyperplane rounding \cite{GoemansW95}, implemented via a random Gaussian vector. (The specific use of a Gaussian vector will actually be important in the following.) Let
$$ \bZ = (Z_1, Z_2, \ldots, Z_n) $$
be a centered multivariate Gaussian with an identity covariance matrix (i.e., $Z_1, Z_2,\ldots,Z_n$ are independent standard Gaussians). Then we define
$$ G_i = \langle \bv_i, \bZ \rangle $$
for each $i \in [n]$. Now, the random variables $(G_1,\ldots,G_n)$ are correlated Gaussians with covariance matrix $\bSigma = (\sigma_{ij})_{i,j}$; $\sigma_{ij} = \Cov[G_i,G_j] = \bv_i \cdot \bv_j$.
In analogy to the Goemans-Williamson rounding, we define a random set
$$ R = \{ i \in [n]: G_i = \langle \bv_i, \bZ \rangle \geq 0 \} $$
and then
$$ F(\bSigma) \coloneqq \E[f(R)].$$
Each element appears in $R$ with probability $1/2$, with correlations determined by the matrix $\bSigma$. It is easy to see that the distribution of $R$ depends only on $\bSigma$, since this determines the vectors $(\bv_1,\ldots,\bv_n)$ up to an isometry.

More generally, we consider a distribution where elements appear with prescribed probabilities, as follows: For $\bx \in [0, 1]^n$ and $\bSigma \in \cE_n$, let
$$ R = \{ i \in [n]: G_i = \langle \bv_i, \bZ \rangle \geq \Phi^{-1}(1 - x_i) \},$$
where $\langle \bv_i, \bv_j \rangle = \sigma_{ij}$, $\bZ \sim N(0,I_n)$. This is a random set where each $i \in [n]$ appears with probability $x_i$, with correlations again determined by $\bSigma$.
Then we define
$$ F(\bx, \bSigma) \coloneqq \E[f(R)].$$
We call this the {\em semidefinite extension} of $f$.
In particular, $\bSigma = I$ corresponds to orthonormal vectors $\bv_1,\ldots,\bv_n$, and Gaussian rounding in this case results in independently sampled elements, each element $i$ appearing in $R$ with probability $x_i$.
In particular, $\bx \mapsto F(\bx, I)$ is the well-studied multilinear extension of $f$, which we also denote by $F(\bx)$.
Furthermore, the reader can check that for the all-ones matrix $J$, $\bx \mapsto F(\bx, J)$ is the Lov\'asz extension of $f$. (We don't use the Lov\'asz extension in this paper.)

Recall \Cref{submodular-on-p-biased-set-approximates-maximum} which, using the notation of the multilinear extension $F(\bx)$, says that $F(p \1) \geq p(1-p) \max_{S \subseteq [n]} f(S)$.
Also, related to this, we have the following comparison.

\begin{lemma}\label{submodular-on-p-biased-set-approximates-mean}
If $f:2^{[n]} \to \R_{\geq 0}$ is nonnegative submodular, $F:[0,1]^n \to \R_{\geq 0}$ its multilinear extension, and $p, q \in (0, 1)$, then
\begin{equation*}
    F(p \1) \geq \min\left\{\frac{p}{q}, \frac{1 - p}{1-q}\right\} F\left(q \1\right).
\end{equation*}
\end{lemma}
For $\bSigma = I$, we obtain $F(I) \geq \frac{1}{4} \OPT$ by \Cref{submodular-on-p-biased-set-approximates-maximum} with $p = 1 / 2$. However, we also obtain a more robust result: If the covariance matrix is $\epsilon$-{\em close} to the identity, then the expected value is at least $(\frac{1}{4} - \epsilon) \OPT$, or at least $(1 - O(\sqrt{\epsilon})) F(I)$. This is based on a stochastic comparison lemma which appears in the literature in various forms  \cite{HoudrePS98,Muller01,Voigtlander21}.
We include a self-contained proof in Appendix~\ref{app:proof-of-submod-comparison}.

\begin{theorem}
\label{thm:submod-comparison}
Let $f : 2^{[n]} \to \R_{\geq 0}$ be nonnegative submodular, and $F:[0,1]^n \times \cE_n \to \R_{\geq 0}$ its semidefinite extension (defined above).
If $\bSigma, \bSigma' \in \cE_n$ are covariance matrices such that $\sigma_{ij} \leq \sigma'_{ij}$ for all $i \neq j$ and $\bx \in [0, 1]^n$, then
$$ F(\bx, \bSigma) \geq F(\bx, \bSigma').$$
\end{theorem}

In other words, lower pairwise covariances imply a higher submodular expectation. This is in line with the intuition that positively correlated elements can cancel out each other's value in a submodular function. However, we note that this intuition is not strictly speaking correct: Bounded pairwise correlations or even pairwise independence are {\em not enough} for a comparison of expectations of submodular functions \cite{ChekuriUV26}. Our intuition is that because a centered multivariate Gaussian is {\em completely determined} by its covariance matrix, the knowledge of covariance coefficients $\sigma_{ij}$ is more than just information about pairwise correlations; the multivariate Gaussian has a rigid structure determined by the covariance matrix which allows our method to work.
\jv{Modified the formulation here, based on a conversation with Richard.}

A consequence of Theorem~\ref{thm:submod-comparison} is that a rounding applied to almost-orthogonal vectors can be compared to a covariance matrix $\bSigma'$ where $\sigma'_{ii} = 1$ and $\sigma'_{ij} = \epsilon > 0$ for $i \neq j$. The uniform structure of $\bSigma'$ makes it easier to analyze its performance via an explicit choice of correlated Gaussian variables. This leads to the following theorem (note that we only need an upper bound on the dot products rather than being close to $0$).

\begin{theorem}
\label{thm:gaussian-rounding}
Let $n, k \geq 1$ be integers, $V = [n]$, and $0 < \epsilon < 1/2$. 
Fix vectors $\bv_1,\ldots, \bv_n \in \R^k$ satisfying $\| \bv_i \| = 1$ and 
$$ \langle \bv_i, \bv_j \rangle \leq \epsilon $$
for $i \ne j$.
Let $\bZ = (Z_1,\ldots,Z_k)$ be a $k$-variate standard Gaussian (with independent coordinates), and let $T$ be another independent random variable of mean $0$ and variance at most $\epsilon$. Let 
$$ R_\bZ(T) = \{ i \in V: \langle \bv_i, \bZ \rangle \geq T \}.$$
Then for every nonnegative submodular function $f : 2^V \to \R$,
$$ \E[f(R_\bZ(T))]  \geq \left( \frac{1}{4} - \epsilon \right) \max_{S \subseteq V} f(S).$$
\end{theorem}

\subsection{Discretization and derandomization in logarithmic dimension}
\label{sec:derandomization}

The second component of the solution is to derandomize the result given by Theorem~\ref{thm:gaussian-rounding}. This is possible due to the fact that a logarithmic dimension is sufficient to construct a near-orthogonal set of $n$ vectors.

\begin{lemma}
\label{lem:near-orthogonal}
Let $n \geq 1$ be an integer, $\epsilon > 0$ and let $k \geq \frac{4}{\epsilon^2} \ln n$ be a positive integer. Then there exist vectors $\bv_1, \ldots, \bv_n \in \R^k$ such that $\|\bv_i\| = 1$ and for $i \ne j$,
$$ |\langle \bv_i, \bv_j \rangle| \leq \epsilon. $$
\end{lemma}

This lemma can be proved directly by either a randomized or deterministic construction. In \Cref{sec:near-orthogonal-randomized} we prove this lemma with a randomized argument, while in Appendix~\ref{app:greedy-near-orthogonal} we give a fully deterministic construction of such vectors. Alternatively, one can view this lemma as a Johnson-Lindenstrauss projection of truly orthonormal vectors from $\R^n$ to $\R^k$. 
Given such a configuration of vectors and the associated Gram matrix $\bSigma$, we know Gaussian rounding gives a random set satisfying $\E[f(R)] \geq (\frac{1}{4} - \epsilon) \OPT$ for a nonnegative submodular $f$, by Theorem~\ref{thm:gaussian-rounding}. It remains to derandomize this construction by identifying a polynomial-size family of sets $R_1,\ldots,R_t$ such that for every submodular function, one of them achieves a value close to the expectation. The key technical lemma here is the following.

\begin{lemma}
\label{lem:discretization}
Let $f:2^{[n]} \to \R_{\geq 0}$ be nonnegative submodular, and let $M = \max_{S \subseteq [n]} f(S)$. Fix $\bz \in \R^k$ and suppose a random vector $\bY = (Y_1,\ldots,Y_k)$ with independent coordinates $Y_1, \dots, Y_k$ satisfies
$\E[\bY] = \bz$, $\E[(Y_j-z_j)^2] \leq \delta$ for all $1 \leq j \leq k$.
Let $\tau$ be a random variable independent of $\bY$, with a continuously differentiable density $p_\tau$.
As above, define $R_\bz(\tau) = \{ i \in [n]: \langle \bv_i, \bz \rangle \geq \tau \}$ for unit vectors $\bv_1, \dots, \bv_n \in \R^k$. Then
\begin{equation*}
    \E_{\bY, \tau}[f(R_\bY(\tau))] \geq \E_\tau[f(R_\bz(\tau))] - M \|p_\tau'\|_\infty \delta.
\end{equation*}
\end{lemma}
The way we use this lemma is that we sample a standard $k$-variate Gaussian vector $\bZ$;
we condition on $\bZ = \bz$ in the following.
We choose $\tau$ as another independent Gaussian of variance $O(\epsilon)$.
We replace $\bz$ by a nearby vector $\bY$ chosen on a grid of subdivision size $O(\epsilon)$, so that $\E[\bY \mid \bZ = \bz] = \bz$.
\Cref{lem:discretization} gives us a bound on how much we lose by replacing $\bz$ by $\bY$.

Finally, we uncondition and consider the expectation over $\bZ$.
This induces a distribution of $\bY$ over the same discrete grid.
We can enumerate all possible outcomes for $\bY$ (in a ball $\|\by\| \leq r$ with $r$ chosen such that $\IP[\|\bY\| > r] = O(\epsilon)$), and the relevant values of $\tau$.
It can be verified that the size of the resulting discretized family of sets $R_\bY(\tau)$ is $n^{\tilde{O}(1/\epsilon^{2})}$.
%\ru{We could say that Chat GPT proved a bound of the form $|\cF| \leq n^{O(\epsilon^{-2})}$, but we chose not to present it for simplicity.}
%\jv{I think for the discussion here, we can just say $n^{\tilde{O}(1/\epsilon^{2})}$, and discuss the details later.}
Since a distribution supported on these sets achieves a value close to the expectation under Gaussian rounding (by $\IP[\|\bY\| > r] = O(\epsilon)$ and \Cref{lem:discretization}), and the expectation under Gaussian rounding is bounded below by $\frac{1}{4} M$ up to a small error (\Cref{thm:gaussian-rounding}), we conclude that one of these sets always achieves a value close to $\frac{1}{4} M$.

\section{Uniform convergence for submodular functions}
\label{sec:uniform-convergence}

Uniform convergence is a fundamental concept in machine learning. Suppose we have a class of functions $\cC$ on a given domain (in our context $2^{[n]}$) and a distribution $\cU$ over the domain (in our context the uniform distribution). For $f \in \cC$, denote $Q(f) = \E_{S \sim \cU}[f(S)]$, and $\hat{Q}_t(f) = \frac{1}{t} \sum_{i=1}^{t} f(S_i)$ where $\sigma = (S_1,\ldots,S_t)$ is a random sample of $t$ independent sets from $\cU$. A central question is whether as the sample $\sigma$ grows, $\hat{Q}_t(f)$ converges to $Q(f)$ {\em uniformly for all functions} $f \in \cC$. This is known as {\em uniform convergence}. Further, {\em polynomial uniform convergence} means that there is a polynomial $p(n, 1/\epsilon, 1/\delta)$ such that for $t \geq p(n, 1/\epsilon, 1/\delta)$:
$$ \Pr[ \exists f \in \cC : | \hat{Q}_t(f) - Q(f) |  > \epsilon \, Q(f)] < \delta.$$
We also consider a weaker notion, with arbitrary dependence on $\delta$ and $\epsilon$. We say $\cC$ admits {\em weak polynomial uniform convergence} if for every $\delta, \epsilon>0$ there is a polynomial $p_{\delta,\epsilon}(n)$ such that for $t \geq p_{\delta,\epsilon}(n)$,
$$ \Pr[ \exists f \in \cC : | \hat{Q}_t(f) - Q(f) |  > \epsilon \, Q(f)] < \delta.$$
\newcommand{\submodular}{\mathrm{sm}}
Theorem~\ref{thm:no-uniform-convergence} shows that even weak polynomial uniform convergence does not happen for the collection $\cC_{\submodular}$ of nonnegative submodular functions:
there are fixed $\delta = 1/2$ and $\epsilon = 1/4$ such that for any polynomial $p(n)$ and sufficiently large $n$,
$$ \Pr[ \exists f \in \cC_{\submodular} : | \hat{Q}_{p(n)}(f) - Q(f) |  > \epsilon \, Q(f)] > \delta.$$
In contrast, we claim that Theorem~\ref{thm:main-exp} shows that for every fixed $\delta,\epsilon>0$, there is a polynomial $p(n)$ and a distribution $\cD$ such that if we sample $t \geq p(n)$ independent samples $S_1,\ldots,S_t \sim \cD$, then
\begin{equation}\label{uniform-convergence-of-mean}
    \Pr\left[ \exists f \in \cC_{\submodular} : \Big| \frac{1}{t} \sum_{i=1}^{t} f(S_i)  - Q(f) \Big|  > \epsilon \, Q(f) \right] \leq \delta.
\end{equation}
Indeed, take $\cD = \cD_{n, \epsilon / 2}$ from \Cref{thm:main-exp}, which has support of size $m = n^{O(1 / \epsilon^4 \log(1 / \epsilon))}$, and take
\begin{equation*}
    t = \left\lceil \frac{32}{\epsilon^2} \left(m \ln(2) + \ln\left(\frac{1}{\delta}\right)\right)\right\rceil = \poly(n)
\end{equation*}
independent samples $S_1, \dots, S_t \sim \cD_{n, \epsilon / 2}$.
\newcommand{\TV}{\mathrm{TV}}
Then we can obtain a bound with high probability on the total variation distance, denoted $d_{\TV}$, of $\frac{1}{t} \sum_{i=1}^t \delta_{S_i}$ and $\cD_{n, \epsilon / 2}$.
By Chernoff's inequality and a union bound
\begin{equation*}
    \IP\left[d_{\TV}\Big(\frac{1}{t}\sum_{i=1}^t \delta_{S_i}, \cD_{n, \epsilon / 2}\Big) \geq \frac{\epsilon}{8}\right] \leq 2^m e^{-t \epsilon^2 / 32} \leq \delta.
\end{equation*}
So the event that $d_{\TV}\Big(\frac{1}{t}\sum_{i=1}^t \delta_{S_i}, \cD_{n, \epsilon / 2}\Big) < \frac{\epsilon}{8}$ holds with probability at least $1 - \delta$.
On that event, any nonnegative $f : 2^{[n]} \to \IR_{\geq 0}$ satisfies
\begin{equation*}
    \Big| \frac{1}{t} \sum_{i=1}^{t} f(S_i)  - \IE_{S \sim \cD_{n, \epsilon / 2}}[f(S)] \Big| \leq \frac{\epsilon}{8} \max_{S \subseteq [n]} f(S),
\end{equation*}
and, for $f$ nonnegative submodular, the right-hand side is at most $\frac{\epsilon}{2} Q(f)$ by \Cref{submodular-on-p-biased-set-approximates-maximum} with $p = 1 / 2$.
Hence, for nonnegative submodular $f : 2^{[n]} \to \IR_{\geq 0}$ and using the properties of $\cD_{n, \epsilon / 2}$ from \Cref{thm:main-exp}, we indeed get
\begin{equation*}
    \Big| \frac{1}{t} \sum_{i=1}^{t} f(S_i)  - Q(f) \Big|
    \leq \Big| \frac{1}{t} \sum_{i=1}^{t} f(S_i)  - \IE_{S \sim \cD_{n, \epsilon / 2}}[f(S)] \Big| + \Big|\IE_{S \sim \cD_{n, \epsilon / 2}}[f(S)] - \IE_{S \sim \cU}[f(S)]\Big|
    \leq \epsilon Q(f).
\end{equation*}
Therefore, \Cref{uniform-convergence-of-mean} follows.

\section{Proofs of the main results}
\label{sec:proofs}

Here we present detailed proofs of our main results.
First, we note some general properties of the multilinear extension $F : [0, 1]^n \to \R_{\geq 0}$ of a nonnegative submodular function $f : 2^{[n]} \to \R_{\geq 0}$.
For $\bx \in [0, 1]^n$, and $i, j \in [n]$,
\begin{equation*}
    \partial_i \partial_j F(\bx) =
    \begin{cases}
    \IE [f(A \cup \{i, j\}) - f(A \cup \{i\}) - f(A \cup \{j\}) + f(A)] & i \neq j,\\
    0 & i = j,
    \end{cases}
\end{equation*}
where the expectation is over $A \subseteq [n] \setminus \{i, j\}$ containing each $k \in [n] \setminus \{i, j\}$ independently with probability $x_k$.
By submodularity of $f$, we conclude that
\begin{equation}\label{F-is-submodular}
    \partial_i \partial_j F \leq 0 \quad \text{and} \quad \|\partial_i \partial_j F\|_\infty \leq
    2 \max_{S \subseteq [n]} f(S).
\end{equation}
Consequently, for any $\bx \in [0, 1]^n$ and $\bv \in \IR_{\geq 0}^n$, we have
\begin{equation*}
    \partial_\bv \partial_\bv F(\bx) = \sum_{i, j = 1}^n v_i v_j \partial_i \partial_j F(\bx) \leq 0.
\end{equation*}
This means that $F$ is concave on any line segment connecting points $\bx, \by \in [0, 1]^n$ for which $x_i \leq y_i$ for all $i \in [n]$.

From these properties, Lemma~\ref{submodular-on-p-biased-set-approximates-maximum} and Lemma~\ref{submodular-on-p-biased-set-approximates-mean} follow quite easily; we defer the proofs to Appendix~\ref{app:proofs}.
A key result that we use is Theorem~\ref{thm:submod-comparison}: a stochastic comparison of submodular functions under Gaussian rounding for two comparable matrices $\bSigma, \bSigma'$.
This result was discovered in various forms in the literature \cite{HoudrePS98,Muller01,Voigtlander21}.
For completeness, we present a proof in Appendix~\ref{app:proofs}.

\subsection{Analysis of Gaussian rounding}

Using these tools, we prove Theorem~\ref{thm:gaussian-rounding}, which is an estimate on the expected value achieved by Gaussian rounding for a near-orthogonal configuration of vectors. 

\begin{proof}[Proof of \Cref{thm:gaussian-rounding}]
    Observe that $G_i = \langle \bv_i, \bZ \rangle$ ($i=1, \dots, n$) are jointly Gaussian with covariance matrix $\bSigma$ satisfying
    \begin{equation*}
        \bSigma_{ij} = \Cov[ G_i, G_j] = \langle \bv_i, \bv_j\rangle.
    \end{equation*}
    Hence, $\sigma_{ii} = 1$ and $\sigma_{ij} \leq \epsilon$ for all $i \neq j$. We can write $\E[f(R_\bZ(t))] = F(\bx, \bSigma)$, where $\bx = (1-\Phi(t)) \1 \in \R^n$.
    Let $(Y_1, \dots, Y_n, Y_0)$ be an $(n+1)$-variate standard Gaussian independent of $T$.
    Then the joint Gaussian $(\sqrt{1-\epsilon} Y_i + \sqrt{\epsilon} Y_0)_{i=1, \dots, n}$ has covariance matrix $\bSigma' = (1-\epsilon) I + \epsilon J$.
    Therefore, by \Cref{thm:submod-comparison},
    \begin{equation*}
        \E[f(R_\bZ(t))] = F(\bx, \bSigma) \geq F(\bx, \bSigma') = \E[f(\{i : \sqrt{1-\epsilon} Y_i + \sqrt{\epsilon} Y_0 \geq t\})] = \E_Y[F(p(t, Y_0) \1)],
    \end{equation*}
    where $p(t, y) = 1 - \Phi\left(\frac{t - \sqrt{\epsilon} y}{\sqrt{1-\epsilon}}\right)$.
    Consequently, by \Cref{submodular-on-p-biased-set-approximates-maximum}
    \begin{equation*}
        \E[f(R_\bZ(T))] \geq \E[F(p(T, Y_0) \1)] \geq \E[p(T, Y_0) (1-p(T, Y_0))]\max_{S \subseteq [n]} f(S).
    \end{equation*}
    Let $U = \frac{T - \sqrt{\epsilon} Y_0}{\sqrt{1 - \epsilon}}$, so $p(T, Y_0) = 1 - \Phi(U)$. We have $\E[U] = 0$ and $\E[U^2] \leq \frac{2\epsilon}{1 - \epsilon}$. Then we obtain
    \begin{equation*}
        \E[\Phi(U)  (1 - \Phi(U))] = \frac{1}{4} - \E\left[\left(\Phi(U) - \frac{1}{2}\right)^2\right] \geq \frac{1}{4} - \|\Phi'\|_\infty^2 \E[U^2] \geq \frac{1}{4} - \frac{\epsilon}{\pi(1-\epsilon)} \geq \frac{1}{4} - \epsilon
    \end{equation*}
    for $\epsilon < \frac{1}{2} < 1 - \frac{1}{\pi}$.
    Consequently, we get the required bound
    \begin{equation*}
        \E[f(R_\bZ(T))] \geq \E[p(T, Y_0) (1-p(T, Y_0))]\max_{S \subseteq [n]} f(S) \geq \left(\frac{1}{4} - \epsilon\right)\max_{S \subseteq [n]} f(S). \qedhere
    \end{equation*}
\end{proof}

\paragraph{Comparison to the expectation of $f$.}
We also derive a related estimate, which instead of the maximum is with respect to the expectation of $f$.

In the proof above, we showed that
$$ \E[f(R_\bZ(T))] \geq \E[F(p(T,Y_0) \1)].$$
Using \Cref{submodular-on-p-biased-set-approximates-mean} with $q = 1 / 2$ instead of \Cref{submodular-on-p-biased-set-approximates-maximum}, we get
$$ F(p(T,Y_0) \1) \geq 2 \min\{p(T, Y_0), 1 - p(T, Y_0)\} F\left(\frac{1}{2}\1\right) = \left(1 - 2|p(T,Y_0) - 1/2|\right) F\left(\frac{1}{2}\1\right).$$
Using the properties of $T$ and $Y_0$, with $U = \frac{T - \sqrt{\epsilon} Y_0}{\sqrt{1-\epsilon}}$ as above,
\begin{equation*}
    \E\left[\left|p(T, Y_0) - \frac{1}{2}\right|\right] = \E\left[\left|\Phi(U) - \frac{1}{2}\right|\right] \leq \|\Phi'\|_\infty \E[|U|] \leq \sqrt{\frac{1}{2\pi}} \sqrt{\frac{2\epsilon}{1-\epsilon}} \leq \sqrt{\epsilon}
\end{equation*}
since $\epsilon < \frac{1}{2} < 1 - \frac{1}{\pi}$.
Hence,
\begin{equation*}
    \E[f(R_\bZ(T))] \geq \E[1 - 2|p(T,Y_0) - 1/2|] F\left(\frac{1}{2}\1\right) \geq (1 - 2 \sqrt{\epsilon}) F\left(\frac{1}{2}\1\right).
\end{equation*}
We summarize this as follows.

\begin{corollary}\label{gaussian-rounding-vs-average}
    Under the assumptions of \Cref{thm:gaussian-rounding}, we get
    \begin{equation*}
        \E[f(R_\bZ(T))] \geq \left(1 - 2\sqrt{\epsilon}\right) \E_{S \subseteq [n]}[f(S)].
    \end{equation*}
\end{corollary}

Here and in the following $\E_{S \subseteq [n]}$ stands for an expectation over a uniformly random subset $S$ of $[n]$.

\subsection{Construction of nearly orthogonal vectors}\label{sec:near-orthogonal-randomized}

Next, we give a proof of \Cref{lem:near-orthogonal}, a construction of near-orthogonal vectors.

\begin{proof}[Proof of \Cref{lem:near-orthogonal}]
    Let us define $\bX_i = (X_{i1}, \ldots, X_{ik})$, where the $X_{ij}$ are independent and uniformly random from $\{1, -1\}$. We have $\| \bX_i \| = \sqrt{k}$, and
    $$ \langle \bX_i, \bX_j \rangle = \sum_{\ell=1}^{k} X_{i\ell} X_{j\ell}. $$
    The products $X_{i\ell} X_{j\ell}$ (for fixed $i \neq j$) are also independent uniform $\pm 1$, and so by Hoeffding's inequality,
    $$ \Pr\left[\left|\sum_{\ell=1}^{k} X_{i\ell} X_{j\ell}\right| > \epsilon k \right] < 2e^{-\epsilon^2 k / 2}.$$
    For $k \geq \frac{4}{\epsilon^2} \ln n$, we get $\Pr[|\langle \bX_i, \bX_j \rangle| > \epsilon k] < \frac{2}{n^2}$ and hence there is a choice $\bX_i = \bx_i$ such that $|\langle \bx_i, \bx_j \rangle| \leq \epsilon k$ for all $i \neq j$ by a union bound over the $\binom{n}{2} < \frac{n^2}{2}$ events. Finally, we set $\bv_i = \frac{1}{\sqrt{k}} \bx_i$ to obtain the vectors claimed by the lemma.
\end{proof}

We note that this construction can also be performed deterministically by a greedy algorithm, see Appendix~\ref{app:greedy-near-orthogonal}.

We can carry out a similar proof with $\bX_i$ having $p$-biased entries in $\{0, 1\}$ instead of in $\{-1, 1\}$. This introduces the slight technical annoyance that then $\|\bX_i\|^2 \approx p k$ instead of the previous exact equality $\|\bX_i\|^2 = k$, which means we have to rescale $\bX_i$ by a random scalar to obtain a unit vector $\bv_i = \frac{1}{\|\bX_i\|} \bX_i$.
The advantage of this construction is that it yields vectors $\bv_i$ with nonnegative entries. Since we will need that for proving \Cref{thm:main-exp}, let us record this new lemma here.

\begin{lemma}\label{lem:near-orthogonal-in-positive-quadrant}
    Let $n \geq 2$ be an integer, $\epsilon \in (0, 1)$ and let $k \geq \frac{100}{\epsilon^2} \ln n$ be a positive integer.
    Then there exist vectors $\bv_1, \ldots, \bv_n \in \R^k$ with nonnegative entries such that $\|\bv_i\| = 1$ and for $i \ne j$,
    $$0 \leq \langle \bv_i, \bv_j \rangle \leq \epsilon. $$    
\end{lemma}
\begin{proof}
    Sample independent subsets $A_1, \dots, A_n \subseteq [k]$, such that each $A_i$ contains every $j \in [k]$ independently with probability $p = \epsilon / 4 \in (0, 1 / 4)$. We have $\E[|A_i|] = pk$ and $\E[|A_i \cap A_j|] = p^2 k$ for $i \neq j$.
    By Chernoff's bound,
    \begin{alignat*}{2}
        \IP\left[|A_i| < \frac{1}{2} pk\right]
        &\leq e^{-pk/8} \leq \frac{1}{2n}
        &&\qquad (i \in [n]),\\
        \IP\left[|A_i \cap A_j| > 2p^2k\right]
        &\leq e^{-p^2k/3} \leq \frac{1}{n^2}
        &&\qquad (i,j \in [n],\ i \neq j),
    \end{alignat*}
    where we used that
    \begin{equation*}
        k \geq \frac{100 \ln n}{\epsilon^2} > \frac{3 \ln(n^2)}{p^2} \geq \frac{8 \ln(2n)}{p}
    \end{equation*}
    by $p = \epsilon / 4 < \frac{3}{8}$ and $n \geq 2$.
    So, the event that $|A_i| \geq \frac{1}{2} pk > 0$ for all $i$ and $|A_i \cap A_j| \leq 2 p^2 k$ for all $i \neq j$ holds with probability at least
    \begin{equation*}
        1 - n \cdot \frac{1}{2n} - \binom{n}{2} \frac{1}{n^2} > 0.
    \end{equation*}
    Condition on this event and let $\bv_i \in \IR_{\geq 0}^k$ be the normalized indicator vector of $A_i$, $\bv_i = \frac{1}{\sqrt{|A_i|}} \1_{A_i}$.
    On this event, all $i \neq j$ indeed satisfy
    \begin{equation*}
        \langle \bv_i, \bv_j\rangle = \frac{|A_i \cap A_j|}{\sqrt{|A_i| |A_j|}} \leq 4 p = \epsilon.
    \end{equation*}
    The bound $\langle \bv_i, \bv_j\rangle \geq 0$ follows from nonnegativity of $\bv_i$ and $\bv_j$.
\end{proof}

\subsection{Discretization of the sampling space}

Here we address the second main step in the proof: reduction to a polynomial number of possible solutions. Recall \Cref{lem:discretization}  which says that instead of a general vector $\bz \in \R^k$, we can consider a nearby random vector $\bY$, and the expected outcome does not change significantly.

\begin{proof}[Proof of \Cref{lem:discretization}]
    Without loss of generality, assume that $\langle \bv_1, \bz \rangle \geq \langle \bv_2, \bz \rangle \geq \cdots \geq \langle \bv_n, \bz \rangle$, so $R_\bz(\tau)$ is supported on the sets $[\ell] = \{1,\ldots,\ell\}$ for $0 \leq \ell \leq n$.
    We define $w_\ell \coloneqq f([\ell]) - f([\ell-1])$ for $1 \leq \ell \leq n$. The submodularity of $f$ yields
    \begin{equation}\label{greedy-inclusion-submodularity-inequality}
        f(A) - f(\emptyset) \geq \sum_{i \in A} w_i
    \end{equation}
    with equality when $A = [\ell]$ for some $0 \leq \ell \leq n$.
    Then
    \begin{equation*}
        \sum_{i \in [n]} |w_i|
        = 2\sum_{i : w_i \geq 0} w_i - \sum_{i \in [n]} w_i
        \leq 2f(\{i : w_i \geq 0\}) - 2 f(\emptyset) - (f([n]) - f(\emptyset)) \leq 2M.
    \end{equation*}
    Since $R_\bz(\tau)$ is supported on the sets $[\ell]$ for $0 \leq \ell \leq n$, \Cref{greedy-inclusion-submodularity-inequality} holds with equality for $A = R_\bz(\tau)$ with probability $1$.
    Consequently, for the cumulative distribution function of $\tau$, $P_\tau(t) = \int_{-\infty}^t p_\tau(u) du$, we have
    \begin{equation*}
        \IE_\tau[f(R_\bz(\tau))] - f(\emptyset) = \sum_{i \in [n]} w_i P_\tau(\langle \bv_i, \bz\rangle),
    \end{equation*}
    and similarly,
    \begin{align*}
        \IE_\tau[f(R_\bY(\tau))] - f(\emptyset)
        &\geq \sum_{i \in [n]} w_i P_\tau(\langle \bv_i, \bY \rangle)\\
        &\geq \sum_{i \in [n]} \left( w_i P_\tau(\langle \bv_i, \bz \rangle) + w_i \langle \bv_i, \bY - \bz \rangle p_\tau(\langle \bv_i, \bz \rangle) - \frac{1}{2} |w_i| \langle \bv_i, \bY - \bz \rangle^2 \|p_\tau'\|_\infty \right),
    \end{align*}
    where we used Taylor's theorem (in degree $2$) in the second inequality.
    Let us denote $V_i \coloneqq \langle \bv_i, \bY - \bz \rangle$, so $\IE_\bY [V_i] = \langle \bv_i, \IE_\bY [\bY - \bz] \rangle = 0$ and $\IE_\bY [V_i^2] = \sum_j v_{ij}^2 \IE_\bY [(Y_j - z_j)^2] \leq \delta \sum_j v_{ij}^2 = \delta$.
    Applying $\IE_\bY$ to both sides above, we get
    \begin{align*}
        \IE_{\bY, \tau}[f(R_\bY(\tau))] - f(\emptyset)
        &\geq \IE_\tau[f(R_\bz(\tau))] - f(\emptyset) - \frac{1}{2} \sum_{i \in [n]} |w_i| \|p_\tau'\|_\infty \delta\\
        &\geq \IE_\tau[f(R_\bz(\tau))] - f(\emptyset) - M \|p_\tau'\|_\infty \delta.
    \end{align*}
    This proves the desired lower bound on $\IE_{\bY, \tau}[f(R_\bY(\tau))]$.
\end{proof}

In order to prove \Cref{thm:main}, we employ Lemma~\ref{lem:discretization} to reduce the space of all possible vectors $\bz$ to a discrete grid with parameters that make the discretization error suitably small. The last part of the proof is a counting argument to bound the number of possible vectors on the grid.

\begin{proof}[Proof of \Cref{thm:main}]
    Without loss of generality, we assume that $0 < \epsilon < 1 / 4$.
    Set $\eta \coloneqq \epsilon / 3$.
    \Cref{lem:near-orthogonal} implies that for $k \coloneqq \lceil 4\eta^{-2} \ln n \rceil > 0$, there are unit vectors $\bv_1, \dots, \bv_n \in \R^k$ satisfying $\langle \bv_i, \bv_j \rangle \leq \eta$ for all $i \neq j$.
    As before, set
    \begin{equation*}
        R_\bz(t) = \{i \in [n] : \langle \bv_i, \bz \rangle \geq t\}
    \end{equation*}
    for $\bz \in \R^k$ and $t \in \R$.
    Sample $\bZ = (Z_1, \dots, Z_k) \sim N(0, I_k)$ and $\tau \sim N(0, \eta)$ independently.
    Conditioned on $\bZ = \bz$, we round each coordinate $z_j$ independently to a nearby (random) $Y_j \in \eta\IZ$ such that $\E[Y_j \mid \bZ=\bz] = z_j$, $|Y_j - z_j| \leq \eta$, and denote  $\bY = (Y_1, \dots, Y_k)$.
    So, $\bY$ conditioned on $\bz$ meets the assumptions of \Cref{lem:discretization}:
    \begin{equation}\label{assumptions-for-subdivision-met}
        \E[\bY \mid \bZ=\bz] = \bz \quad \text{and} \quad \E[(Y_j - z_j)^2 \mid \bZ=\bz] \leq \eta^2.
    \end{equation}
    Now let $f : 2^{[n]} \to \IR$ be nonnegative submodular, and denote $M \coloneqq \max_{S \subseteq [n]} f(S)$.
    By Lemma~\ref{lem:discretization}, with $\delta = \eta^2$, we have $\E_{\bY,\tau}[f(R_\bY(\tau)) \mid \bZ=\bz] \geq \E_\tau[f(R_\bz(\tau))] - M \| p'_\tau \|_\infty \eta^2$. Removing the conditioning on $\bZ=\bz$, we have 
    $$\E_{\bY,\tau}[f(R_\bY(\tau))] \geq \E_{\bZ,\tau}[f(R_\bZ(\tau))] - M \| p'_\tau \|_\infty \eta^2,$$
    where we used that $\tau \sim N(0,\eta)$ has differentiable density $p_\tau(t) = \frac{1}{\sqrt{2\pi \eta}} e^{-t^2 / (2\eta)}$.
    In particular,
    \begin{equation*}
    \|p_\tau'\|_\infty = \sup_{t \in \R} \frac{|t|}{\sqrt{2\pi \eta^3}} e^{-t^2 / (2\eta)} = \frac{1}{\sqrt{2 \pi e} \eta} < \frac{1}{\eta}.
    \end{equation*}
    Hence,
    $$\E_{\bY,\tau}[f(R_\bY(\tau))] \geq \E_{\bZ,\tau}[f(R_\bZ(\tau))] - M \eta.$$
    Finally, \Cref{thm:gaussian-rounding} implies $\E_{\bZ,\tau}[f(R_\bZ(\tau))] \geq \left(\frac{1}{4} - \eta\right) M$ by our choice of $\bv_1, \dots, \bv_n$ and $\Var[\tau] = \eta$.
    Combining all these bounds yields
    \begin{align*}
%        \E[f(R_\bY(\tau)) \1_{\{\|\bY\| \leq r\}}] \geq
       \E[f(R_\bY(\tau))]
        &\geq \E[f(R_\bZ(\tau))] - \eta M \geq \left( \frac{1}{4} - 2\eta \right) M.
    \end{align*}
    It remains to count the number of possible solutions generated by $\bY$ and $\tau$.
    We will restrict the solutions to the event $\|\bY\| \leq r$ for large enough $r$,
    which will not affect the expectation significantly.
    The bound $\E[\|\bY\|^2] = \E[\|\bZ\|^2] + \E[\|\bY - \bZ\|^2] \leq 2k$ gives
    \begin{equation*}
        \IP[\|\bY\| > r] = \IP[\|\bY\|^2 > r^2] < \frac{2k}{r^2} = \eta
    \end{equation*}
        for $r \coloneqq \sqrt{2k / \eta}$. We set
    \begin{equation*}
        \cF_{n,\epsilon} \coloneqq \{R_\by(t) : \by \in \eta \IZ^k, \|\by\| \leq r, t \in \IR\}.
    \end{equation*}
    This captures all the solutions obtained as $R_\bY(\tau)$ when $\|\bY\| \leq r$.
    We bound the number of such sets using a volume estimate for $\ell^2$-balls in $\IR^k$ as follows. Note that the open $\ell_\infty$-balls of radius $1 / 2$ centered at the points $\{ \by \in \IZ^k : \|\by\| \leq r / \eta \}$ are disjoint of volume $1$.
    Since they are all contained in the $\ell_2$-ball with center $0$ and radius $\sqrt{k} / 2 + r / \eta \leq 2 r / \eta$, we get
    \begin{equation*}
        \left|\left\{ \by \in \IZ^k : \|\by\| \leq \frac{r}{\eta}\right\}\right|
        \leq \left(\frac{2r}{\eta}\right)^k \int_{\by \in \IR^k} \bone_{\{\|\by\| \leq 1\}} d\by \leq \left(\frac{2r}{\eta}\right)^k \int_{\by \in \IR^k} e^{k (1 - \|\by\|^2) / 2} d\by = \left(\frac{2r}{\eta} \sqrt{\frac{2\pi e}{k}}\right)^k.
    \end{equation*}
    For each fixed $\by$, there are at most $n+1$ relevant choices for $t \in \R$ (determined by the values of $\langle \bv_i, \by \rangle$ for $i=1,\ldots,n$). Therefore, we have
    \begin{equation*}
        |\cF_{n,\epsilon}| \leq (n+1) \left(\frac{2r}{\eta} \sqrt{\frac{2\pi e}{k}}\right)^k
        = (n+1) \left( \frac{4 \sqrt{\pi e}}{\eta^{3 / 2}} \right)^k
        = e^{O(\log n \cdot \epsilon^{-2} \log(\epsilon^{-1}))}
    \end{equation*}
    and
    $$ \max_{S \in \cF_{n,\epsilon}} f(S) \geq \E[f(R_\bY(\tau)) \1_{\{\|\bY\| \leq r\}}]
    \geq \E[f(R_\bY(\tau))] - \Pr[\|\bY\| > r] \cdot M \geq \left( \frac{1}{4} - 3 \eta \right) M $$ 
    as desired.
\end{proof}

\subsection{Comparison to the expectation of $f$}

Let us move now to Theorem~\ref{thm:main-exp}, which gives distributions $\cD_{n,\epsilon}$ of polynomial-size support such that sampling from $\cD_{n,\epsilon}$ approximates the {\em uniform expectation} of every nonnegative submodular function.  The approach is similar to the proof of Theorem~\ref{thm:main}, but it requires some more care in several places. In particular, we need a two-sided bound for $\E_{S \sim \cD_{n,\epsilon}}[f(S)]$ as opposed to the one-sided bound of Theorem~\ref{thm:main}.

An additional tool that we need for upper-bounding $\E_{S \sim \cD_{n,\epsilon}}[f(S)]$ is the following inequality due to Christofides and Vaggelatou (\cite[Theorem 1 (a)]{ChristofidesV}): Let $\bX$ and $\tilde \bX$ be random vectors in $\{0,1\}^n$ with expectation $\E[\bX] = \E[\tilde \bX] = \bx$, where $\bX$ is a monotonic function of independent random variables, while the coordinates of $\tilde \bX$ are independent.  Then any submodular function $f : 2^{[n]} \to \IR$ satisfies $\E[f(\bX)] \leq \E[f(\tilde \bX)]$. We restate this formally in the following lemma and give a short proof based on the FKG correlation inequality.

\begin{lemma}\label{correlation-inequality}
    Let $U_1, \dots, U_k$ be independent random variables, and let $\psi: \R^k \to \{0,1\}^n$ be measurable and monotonic, i.e., $\psi(u_1, \dots, u_k) \geq \psi(u'_1, \dots, u'_k)$ coordinate-wise, whenever $\bu \geq \bu'$ coordinate-wise.
    Define $\bX \coloneqq \psi(U_1, \dots, U_k) \in \{0,1\}^n$ and let $x_i \coloneqq \IP[X_i=1]$.
    Then for any submodular $f : \{0,1\}^{n} \to \IR$ with multilinear extension $F$, we have
    \begin{equation*}
        \E[f(\bX)] \leq F(\bx).
    \end{equation*}
\end{lemma}
\begin{proof}
We proceed by induction on $n$.
The case $n = 0$ is trivial.
Now assume that $n \geq 1$. Write $\bX' = (X_1, \dots, X_{n-1})$ and define
$f_0(\bX') = f(\bX', 0)$, $f_1(\bX') = f(\bX', 1)$, and $d(\bX') = f_1(\bX') - f_0(\bX')$. Then $d(\bX')$ is decreasing in $\bU$ by the submodularity of $f$, while $X_n$ is increasing in $\bU$.
Therefore, the FKG inequality (see, for example \cite[Theorem 6.2.1]{probabilistic-method}) implies that
\begin{align*}
    \E[f(\bX)]
    &= \E[f_0(\bX')] + \E[X_n d(\bX')]\\
    &\leq \E[f_0(\bX')] + \E[X_n] \, \E[d(\bX')]\\
    &= (1-x_n) \, \E[f_0(\bX')] + x_n \, \E[f_1(\bX')].
\end{align*}
The claim now follows by applying the inductive hypothesis to the submodular functions $f_0, f_1$ and the random vector $\bX' = (\psi(\bU)_1, \dots, \psi(\bU)_{n-1})$.
\end{proof}

We are now ready to prove \Cref{thm:main-exp}.
\begin{proof}[Proof of \Cref{thm:main-exp}]
    We start by defining the distribution $\cD_{n,\epsilon}$ and lower-bounding $\E_{S \sim \cD_{n,\epsilon}}[f(S)]$.
    This part of the proof mirrors the proof of \Cref{thm:main}; however, we use a slightly different construction of the vectors $\bv_1,\ldots,\bv_n$. 
    Without loss of generality, we assume that $0 < \epsilon < 1 / 4$. Set $\eta \coloneqq (\epsilon / 3)^2$.
    \Cref{lem:near-orthogonal-in-positive-quadrant} implies that for $k \coloneqq \lceil 100 \eta^{-2} \ln(n)\rceil$, there are unit vectors $\bv_1, \dots, \bv_n \in \R_{\geq 0}^k$ ({\em with nonnegative entries}) satisfying $\langle \bv_i, \bv_j \rangle \leq \eta$ for all $i \neq j$.
    We sample again a standard multivariate Gaussian $\bZ = (Z_1, \dots, Z_k) \sim N(0, I_k)$, and independently, $\tau \sim N(0, \eta)$. Next, we round $\bZ$ to a grid in $\R^k$: Let us introduce an independent random vector $\bU = (U_1, \dots, U_k)$ with independent uniformly distributed entries $U_j \in [-1/2, 1/2]$, and let $Y_j$ be the closest point in $\eta \IZ$ to $Z_j + \eta U_j$ (a tie happens with probability $0$, in which case we break it arbitrarily).
    So $\bY = (Y_1, \dots, Y_k) \in \eta \IZ^k$ has conditionally independent entries given $\bZ=\bz$, and satisfies \Cref{assumptions-for-subdivision-met}.
    We define again
    \begin{equation*}
        R_\by(t) = \{i \in [n] : \langle \bv_i, \by \rangle \geq t\}
    \end{equation*}
    for any $\by \in \R^k$ and $t \in \R$.
    We also again let $r \coloneqq \sqrt{2k / \eta}$, so $\IP[\|\bY\| \leq r] \geq 1 - \eta$, and let
    \begin{equation*}
        \cF \coloneqq \{R_\by(t) : \by \in \eta \IZ^k, \|\by\| \leq r, t \in \IR\}.
    \end{equation*}
    
    Further, we let $\tilde{R}_\bY(\tau)$ be a random set that equals $R_\bY(\tau)$ when $\|\bY\| \leq r$ and that is uniform on $\{\emptyset, [n]\} \subseteq \cF$ conditioned on $\|\bY\| > r$.
    Then $\tilde{R}_\bY(\tau)$ is supported on $\cF \cup \{\emptyset, [n]\} = \cF$.
    We claim that the distribution of $R_\bY(\tau)$ conditioned on $\|\bY\| > r$ is symmetric under taking complements.
    Indeed, this follows from the facts that the distribution of $(\bY, \tau)$ is symmetric with respect to negation, and that $R_{-\bY}(-\tau)$ is the complement of $R_\bY(\tau)$ with probability $1$ (since $\tau$ is continuous).
    Using this and the submodularity inequality $f(A) + f(A^c) \geq f(\emptyset) + f([n])$, we get that the distribution $\cD_{n,\epsilon}$ of $\tilde{R}_\bY(\tau)$ satisfies
    \begin{equation}\label{ub-for-f-eval-at-D-n}
        \E_{S \sim \cD_{n,\epsilon}}[f(S)] = \E[f(\tilde{R}_\bY(\tau))] \leq \E[f(R_\bY(\tau))].
    \end{equation}
    By the same computation as in the proof of \Cref{thm:main}, we obtain the desired bound on the support of $\cD_{n,\epsilon}$:
    \begin{equation*}
        |\supp \cD_{n,\epsilon}| \leq |\cF| \leq \exp(O(\log n \cdot \eta^{-2} \log(\eta^{-1}))) = \exp(O(\log n \cdot \epsilon^{-4} \log(\epsilon^{-1}))).
    \end{equation*}
    For a nonnegative submodular $f$ with $\max_{S \subseteq [n]} f(S) = M$, we get $M \leq 4 \IE_{S \subseteq [n]}[f(S)]$ by \Cref{submodular-on-p-biased-set-approximates-maximum} with $p = 1 / 2$.
    Then
    \begin{equation*}
        \E_{S \sim \cD_{n,\epsilon}}[f(S)]
        \geq \E[f(R_\bZ(\tau))] - 2\eta M
        \geq \left(1 - 2 \sqrt{\eta} - 8 \eta \right)\E_{S \subseteq [n]}[f(S)]
        \geq \left(1 - \epsilon\right)\E_{S \subseteq [n]}[f(S)],
    \end{equation*}
    where the second-to-last inequality follows from \Cref{gaussian-rounding-vs-average}, and the last inequality follows from
    $2 \sqrt{\eta} + 8 \eta \leq 3 \sqrt{\eta} = \epsilon$ by $\eta = (\epsilon / 3)^2 < \frac{1}{144}$.
    This establishes the required lower bound on $\E_{S \sim \cD_{n,\epsilon}}[f(S)]$.
    
    We now turn to the upper bound.
    Instead of using the discretization lemma (\Cref{lem:discretization}), we will directly upper bound $\E[f(R_\bY(\tau))] \leq \IE_{S \subseteq [n]}[f(S)]$.
    This follows by \Cref{correlation-inequality} and the fact that we can write $R_\bY(\tau)$ as an increasing function of independent random variables:
    \begin{equation*}
        (R_\bY(\tau))_i = \1_{\{\left\langle \bv_i, \bY \right\rangle \geq \tau\}} \quad \mbox{ for } i \in [n],
    \end{equation*}
    and $\bY$ is increasing as a function of $(\bZ, \bU)$.
    Consequently, $R_\bY(\tau)$ is increasing as a function of $(\bZ, \bU, -\tau)$.
    Furthermore, since the distribution of $\tau$ is continuous and the joint distribution of $(\bY, \tau)$ is symmetric around $0$, it holds that $\E[1_{\{i \in R_\bY(\tau)\}}] = 1 / 2$ for all $i \in [n]$. 
    Therefore, by \Cref{correlation-inequality}, we indeed get $\E[f(R_\bY(\tau))] \leq \IE_{S \subseteq [n]}[f(S)]$.
    Combining this with \Cref{ub-for-f-eval-at-D-n}, we get
    \begin{equation*}
        \E_{S \sim \cD_{n,\epsilon}}[f(S)] \leq \E[f(R_\bY(\tau))] \leq \E_{S \subseteq [n]} f(S).\qedhere
    \end{equation*}
\end{proof}

\subsection{Counterexample to uniform convergence}

Now we turn to the negative result of Theorem~\ref{thm:no-uniform-convergence}.

\begin{proof}[Proof of Theorem~\ref{thm:no-uniform-convergence}]
Let $\eta = \eta(\delta) \in (0, 1 / 4)$ be a small constant in terms of $\delta$ to be determined.
Let $\cF = \{ A_1, \ldots, A_t \}$ be a family of random subsets of $[n]$, sampled independently from the uniform distribution. Assuming that $t = poly(n)$, it holds with high probability that $|A_i| \in [(\frac{1}{2}-\eta) n, (\frac{1}{2}+\eta) n]$ and $|A_i \cap A_j| < (\frac{1}{4}+\eta) n$ for all $i \neq j$. Let us assume from now on that this is the case. We define the following function $g:[0,1]^n \to \R$:
$$ g(\bx) = \min \left\{ \sum_{i=1}^{n} x_i, \sum_{i=1}^{n} (1-x_i), \min_{1 \leq j \leq t} \left( \left( \frac{3}{4} + \eta \right) n - \sum_{i \in A_j} x_i + \sum_{i \notin A_j} x_i \right) \right\}.$$
It is easy to see that $g$ is nonnegative and continuous.
We claim that $g$ has partial derivatives almost everywhere on each axis-aligned line segment, and each partial derivative $\frac{\partial g}{\partial x_i}$ is non-increasing in the variables $x_1,\ldots,x_n$. By integration and continuity of $g$, it then follows that $g$ restricted to $\{0,1\}^n$ is submodular.

Indeed, if we fix all the variables except for $x_i$, $g(\bx)$ is a minimum of finitely many linear functions in $x_i$, which is differentiable except at finitely many points.
To prove monotonicity of the partial derivatives, define regions $\cA_j$ where $\cA_j$ is the set of points where the minimum is attained by $\left( \frac{3}{4} + \eta \right) n - \sum_{i \in A_j} x_i + \sum_{i \notin A_j} x_i$. We also define $\cA_+, \cA_-$ as the sets of points where the minimum is attained by $\sum_{i=1}^{n} x_i$ or $\sum_{i=1}^{n} (1-x_i)$, respectively. If $\sum_{i=1}^{n} x_i \leq n/2$, $\bx \in \cA_j$ implies $\sum_{i \in A_j} x_i \geq (\frac{3}{8} + \frac{\eta}{2}) n$ and hence $\sum_{i \notin A_j} x_i \leq (\frac{1}{8} - \frac{\eta}{2}) n$. If $\sum_{i=1}^{n} x_i \geq n/2$, $\bx \in \cA_j$ implies $\sum_{i \notin A_j} x_i \leq (\frac{1}{8} - \frac{\eta}{2}) n$ and hence $\sum_{i \in A_j} x_i \geq (\frac{3}{8} + \frac{\eta}{2}) n$. Either way, $\bx \in \cA_j$ implies both $\sum_{i \in A_j} x_i \geq (\frac{3}{8} + \frac{\eta}{2}) n$ and $\sum_{i \notin A_j} x_i \leq (\frac{1}{8} - \frac{\eta}{2}) n$.

The function is linear on each region. Hence, we need to check pair of points such that $\bx \leq \bx'$ and each point is in a different region. Assume that $\bx \leq \bx'$, $\bx \in \cA_j$ and $\bx' \in \cA_{j'}$. Then we have 
$$ \sum_{i \in A_j \setminus A_{j'}} x_i \leq \sum_{i \notin A_{j'}} x_i \leq \sum_{i \notin A_{j'}} x'_i \leq \left(\frac{1}{8} - \frac{\eta}{2} \right) n.$$
On the other hand, we have $\sum_{i \in A_j} x_i \geq (\frac{3}{8} + \frac{\eta}{2}) n$, which implies
$$ \sum_{i \in A_j \cap A_{j'}} x_i = \sum_{i \in A_j} x_i - \sum_{i \in A_j \setminus A_{j'}} x_i 
\geq \left( \frac{3}{8} + \frac{\eta}{2} \right) n - \left( \frac{1}{8} - \frac{\eta}{2} \right) n = \left( \frac{1}{4} + \eta \right) n.$$
This is impossible if $j \neq j'$, because $|A_j \cap A_{j'}| < (\frac{1}{4} + \eta) n$. From here, we conclude two facts: the regions $\cA_j$ are disjoint (by plugging in $\bx = \bx'$), and moreover there is no comparable pair of points $\bx \leq \bx'$ such that $\bx \in \cA_j, \bx' \in \cA_{j'}$, $j \neq j'$. 

Hence, the only three possible transitions between regions for $\bx \leq \bx'$ are described by
\begin{equation*}
    (\bx, \bx') \in \left(A_+ \times A_-\right) \cup \bigcup_j \left( A_+ \times \cA_j\right) \cup \bigcup_j \left(\cA_j \times A_-\right).
\end{equation*}
In all three cases, we have $\frac{\partial g}{\partial x_i}(\bx) \geq \frac{\partial g}{\partial x_i}(\bx')$, by definition of $g$ in the respective regions. 
This concludes the proof that $g$ is submodular on $\{0,1\}^n$.

For every set $A_i \in \cF$, we have $g(\bone_{A_i}) = \min \{ |A_i|, n-|A_i|, (3/4+\eta) n-|A_i| \}$.
Since $|A_i| \geq (\frac{1}{2} - \eta) n$, we have $g(\bone_{A_i}) \leq (\frac{1}{4} + 2\eta) n$. 

Now consider a uniformly random set $R$.
It holds with high probability that $|R| \in [(\frac{1}{2}-\eta) n, (\frac{1}{2}+\eta) n]$ and $|R \cap A_i| \leq (\frac{1}{4} + \eta) n$ and $|R \setminus A_i| = |R|-|R \cap A_i|$.
On this event,
\begin{align*}
    g(\1_R)
    &= \min \Big\{ |R|, n - |R|, \min_{1 \leq i \leq t} \Big(\frac{3}{4} + \eta\Big) n +|R| - 2|R \cap A_i| \Big\}\\
    &\geq \min\Big\{\Big(\frac{1}{2} - \eta\Big) n, \Big(\frac{3}{4} - 2\eta\Big) n\Big\}\\
    &\geq \Big(\frac{1}{2} - \eta\Big) n.
\end{align*}
Therefore,
\begin{equation*}
    \E[g(\bone_R)]
    \geq \Big(\frac{1}{2} - \eta - o(1)\Big) n.
\end{equation*}
Hence the ratio between the maximum over $\cF$ and the uniform expectation is at most $\frac{\frac{1}{4} + 2\eta}{\frac{1}{2} - \eta - o(1)} = \frac{1}{2} + O(\eta) + o(1)$.
Choosing $n$ large enough and a small enough $\eta = \eta(\delta) > 0$, this will be bounded by $\frac{1}{2} + \delta$.
\end{proof}

\bibliographystyle{amsplain}
\bibliography{refs}

\appendix

\section{Proofs of some known or folklore results}
\label{app:proofs}

For completeness, we prove several results, which are not new, used above, either because their proofs might be difficult to extract from the literature or because we give a simpler proof here.

\subsection{Proofs of \Cref{submodular-on-p-biased-set-approximates-maximum,submodular-on-p-biased-set-approximates-mean}}

\begin{proof}[Proof of \Cref{submodular-on-p-biased-set-approximates-maximum}]
    Suppose $f$ attains its maximum at $S$. We use $S^c$ to denote the complement of $S$.
    Concavity of $F$ on the line segment connecting $p \1_{S^c}$ and $p \1_{S^c} + \1_S$ yields
    \begin{equation*}
        F(p \1) \geq (1-p)F(p \1_{S^c}) + p F(p \1_{S^c} + \1_S) \geq p F(p \1_{S^c} + \1_S),
    \end{equation*}
    where we used the nonnegativity of $F$.
    Similarly, concavity of $F$ on the line connecting $\1_S$ and $\1$ yields
    \begin{equation*}
        F(p \1_{S^c} + \1_S) \geq (1-p)F(\1_S) + p F(\1) \geq (1-p) F(\1_S) = (1-p) f(S),
    \end{equation*}
    where we used the nonnegativity of $F$.
    Together, these bounds imply $F(p \1) \geq p F(p \1_{S^c} + \1_S) \geq p(1-p) f(S)$.
\end{proof}
\begin{proof}[Proof of \Cref{submodular-on-p-biased-set-approximates-mean}]
    Without loss of generality, we may assume that $p \leq q$.
    Otherwise, we instead consider the submodular function $S \mapsto f(S^c)$ with $p$ and $q$ replaced by $1 - p$ and $1 - q$.
    By $0 < p \leq q$, we have that $p \1 = \frac{p}{q} (q \1)$ lies on the line segment connecting the origin and $q \1$.
    Therefore, concavity of $F$ on that line yields
    \begin{equation*}
        F(p \1) \geq \left(1 - \frac{p}{q}\right) F({\bf 0}) + \frac{p}{q} F(q \1) \geq \frac{p}{q} F(q \1),
    \end{equation*}
    where we used the nonnegativity of $F$.
\end{proof}

\subsection{Proof of \Cref{thm:submod-comparison}} \label{app:proof-of-submod-comparison}

Here we present a self-contained proof of the stochastic comparison theorem due to \cite{Muller01}.

\begin{claim}\label{multivariate-gaussian-integration-by-parts}
    Suppose $(U, V_1, \dots, V_n)$ is a centered $(n+1)$-variate Gaussian, and let $g : \R^n \to \R$ be continuously differentiable with bounded partial derivatives.
    Then for $\bV = (V_1, \dots, V_n)$,
    \begin{equation*}
        \E[U g(\bV)] = \sum_{j=1}^n \Cov[U, V_j] \, \E[\partial_j g(\bV)].
    \end{equation*}
\end{claim}
\begin{proof}
    First note that $|g(\bV)| \leq |g({\bf 0})| + \|\bV\| \sup_{\bv \in \IR^n} \|\nabla g(\bv)\| = O(1 + \|\bV\|)$, so the expectation on the left-hand side is well-defined.
    For $U \equiv 0$, both sides equal $0$, so the claim holds.
    Now suppose otherwise that $\sigma^2 = \Var[U] > 0$.
    We use the following one-dimensional special case which is known as Gaussian integration by parts: For continuously differentiable $h : \R \to \R$ with $\|h'\|_\infty < \infty$, integration by parts yields
    \begin{equation}\label{gaussian-integration-by-parts}
        \E[U h(U)]
        = \int_{-\infty}^\infty h(u) \cdot \frac{u}{\sqrt{2\pi \sigma^2}} e^{-u^2 / (2 \sigma^2)} du
        = \int_{-\infty}^\infty h'(u) \cdot \sigma^2 \frac{1}{\sqrt{2\pi \sigma^2}} e^{-u^2 / (2 \sigma^2)} du
        = \sigma^2 \E[h'(U)].
    \end{equation}
    Define $\ba = (a_1, \dots, a_n) \in \IR^n$ by $a_j \coloneqq \frac{1}{\sigma^2} \Cov[U, V_j]$.
    Then $U$ and $\bW = \bV - U \ba$ are Gaussians with $\Cov[U, W_j] = \Cov[U, V_j] - a_j \sigma^2 = 0$, hence $U$ and $\bW$ are independent.
    In particular, the conditional distribution of $U$ given $\bW$ is still centered with variance $\sigma^2$. Hence, \Cref{gaussian-integration-by-parts} yields
    \begin{align*}
        \E[U g(\bV) \mid \bW]
        &= \E[U g(\bW + U \ba) \mid \bW]\\
        &= \sigma^2 \E\left[\frac{d}{du} g(\bW + u \ba)\mid_{u = U} \,\middle\vert\, \bW \right] \\
        &= \sigma^2 \E\left[\sum_j a_j \partial_j g(\bW + U \ba) \,\middle\vert\, \bW\right]\\
        &= \sum_j \Cov[U, V_j] \, \E[\partial_j g(\bV) \mid \bW]
    \end{align*}
    by \Cref{gaussian-integration-by-parts}. Taking an expectation over $\bW$ yields the claimed equality.
\end{proof}

\begin{proof}[Proof of \Cref{thm:submod-comparison}]
First, we define approximations of $f(\{i : z_i \geq \Phi^{-1}(1 - x_i)\})$, where the sharp cut-offs $1_{\{z_i \geq \Phi^{-1}(1 - x_i)\}}$ are replaced with smooth approximations as follows:
Let $\theta : \R \to [0, 1]$ be smooth and non-decreasing with $\theta(s) = 0$ for $s \leq -1$ and $\theta(s) = 1$ for $s \geq 0$.
For $\epsilon > 0$, we define the smooth functions $q_{i, \epsilon}(z) = \theta\left(\frac{\Phi(z) - (1-x_i)}{\epsilon}\right), z \in \IR$ and $h_\epsilon(\bz) = F(q_{1, \epsilon}(z_1), \dots, q_{n, \epsilon}(z_n)), \bz \in \IR^n$.
Then $\lim_{\epsilon \downarrow 0} q_{i, \epsilon}(z_i)$ is the indicator of $z_i \geq \Phi^{-1}(1-x_i)$. Hence, by continuity of $F$,
\begin{equation}\label{h-epsilon-approximates-f-of-R}
    \lim_{\epsilon \downarrow 0} h_\epsilon(\bz) = f(\{i : z_i \geq \Phi^{-1}{(1 - x_i)}\}).
\end{equation}
For all $i \neq j$, by \Cref{F-is-submodular},
\begin{equation}\label{h-epsilon-is-submodular}
    \partial_i \partial_j h_\epsilon(\bz) = q_{i, \epsilon}'(z_i) q_{j, \epsilon}'(z_j) \partial_i \partial_j F(q_{1, \epsilon}(z_1), \dots, q_{n, \epsilon}(z_n)) \leq 0.
\end{equation}
Since each $q_{i, \epsilon}$ and $F$ is bounded with bounded first and second partial derivatives, we similarly get that $h_\epsilon$ is bounded with bounded first and second partial derivatives.
Take independent $\bX = (X_1, \dots, X_n) \sim N(0, \bSigma)$ and $\bY = (Y_1, \dots, Y_n) \sim N(0, \bSigma')$, and set $\bZ^{(t)} = \sqrt{1-t} \bX + \sqrt{t} \bY = (Z^{(t)}_1, \dots, Z^{(t)}_n)$ for $t \in [0, 1]$.
Since $h_\epsilon$ is bounded and smooth, we have for $t \in (0, 1)$,
\begin{equation*}
    \frac{d}{dt} \E[h_\epsilon(\bZ^{(t)})]
    = \sum_i \E\left[\partial_i h_\epsilon(\bZ^{(t)}) \left(\frac{Y_i}{2\sqrt{t}} - \frac{X_i}{2\sqrt{1-t}}\right)\right] = \frac{1}{2} \sum_{i, j} (\sigma_{ij}' - \sigma_{ij}) \E\left[\partial_i \partial_j h_\epsilon(\bZ^{(t)})\right],
\end{equation*}
where we used \Cref{multivariate-gaussian-integration-by-parts} and that $\Cov[Y_i, Z^{(t)}_j] = \sigma_{ij}'\sqrt{t}$ and $\Cov[X_i, Z^{(t)}_j] = \sigma_{ij}\sqrt{1-t}$.
Since $\partial_i \partial_j h_\epsilon(\bZ^{(t)}) \leq 0$ for $i \neq j$ by \Cref{h-epsilon-is-submodular} and since $\sigma_{ij} \leq \sigma_{ij}'$ with equality when $i = j$, we conclude that $\frac{d}{dt} \E[h_\epsilon(\bZ^{(t)})] \leq 0$.
Consequently,
\begin{equation*}
    \E[h_\epsilon(\bX)] = \E[h_\epsilon(\bZ^{(0)})] \geq \E[h_\epsilon(\bZ^{(1)})] = \E[h_\epsilon(\bY)].
\end{equation*}
By \Cref{h-epsilon-approximates-f-of-R} and since $\|h_\epsilon\|_\infty \leq \max_{S \subseteq [n]} f(S) < \infty$, the left-hand side converges to $F(\bx, \bSigma)$ and the right-hand side converges to $F(\bx, \bSigma')$ as $\epsilon$ tends to $0$.
Therefore, $F(\bx, \bSigma) \geq F(\bx, \bSigma')$, as required.
\end{proof}

\subsection{Greedy construction of nearly orthogonal vectors}
\label{app:greedy-near-orthogonal}

We show that the nearly orthogonal vectors from
\Cref{lem:near-orthogonal} can be constructed
deterministically by a simple greedy algorithm.
The construction achieves the same dimension
$k=O(\varepsilon^{-2}\log n)$ as the probabilistic
argument, with a running time of
$n^{O(\varepsilon^{-2})}$.

\paragraph{The greedy algorithm.}
Let $n\geq 2$, $\varepsilon\in(0,1)$, and set
\[
k=\left\lceil\frac{4\ln n}{\varepsilon^2}\right\rceil.
\]
Consider the set of normalized sign vectors
\[
\mathcal V=
\left\{
\frac{1}{\sqrt{k}}\bx:
\bx\in\{-1,1\}^k
\right\}.
\]

We construct $\bv_1,\ldots,\bv_n$ sequentially.
At step $i$, having already chosen
$\bv_1,\ldots,\bv_{i-1}$, we enumerate the
vectors in $\mathcal V$ in a fixed order and
select the first vector $\bv_i$ satisfying
\[
|\langle\bv_i,\bv_j\rangle|\leq\varepsilon
\qquad\text{for every }j<i.
\]

The following lemma guarantees that such a vector
always exists, and that the entire procedure runs
in polynomial time for every fixed $\varepsilon>0$.

\begin{lemma}
\label{lem:greedy-near-orthogonal}
For every $n\geq 2$ and $\varepsilon\in(0,1)$,
the greedy algorithm constructs $n$ unit vectors
$\bv_1,\ldots,\bv_n\in\mathbb R^k$, where
$k=\lceil4\varepsilon^{-2}\ln n\rceil$, satisfying
\[
|\langle\bv_i,\bv_j\rangle|\leq\varepsilon
\qquad\text{for all }i\neq j.
\]
Its running time is
$O(n^2k2^k)=n^{O(\varepsilon^{-2})}$.
\end{lemma}

\begin{proof}
Suppose that $\bv_1,\ldots,\bv_{i-1}$
have already been constructed.
We show that at least one vector in $\mathcal V$
satisfies all the required inequalities.

Consider a uniformly random vector
$\bV\in\mathcal V$.
For any previously chosen vector $\bv_j$,
the inner product $\langle\bV,\bv_j\rangle$
has the same distribution as
\[
\frac{1}{k}\sum_{\ell=1}^k X_\ell,
\]
where $X_1,\ldots,X_k$ are independent
uniform random signs.
By Hoeffding's inequality,
\[
\Pr\left[
|\langle\bV,\bv_j\rangle|>\varepsilon
\right]
\leq
2e^{-\varepsilon^2k/2}
\leq \frac{2}{n^2}.
\]
Applying the union bound over all previously
chosen vectors, we obtain
\[
\begin{aligned}
&\Pr\left[
\exists j<i:
|\langle\bV,\bv_j\rangle|>\varepsilon
\right]\\
&\qquad\leq
\sum_{j=1}^{i-1}
\Pr\left[
|\langle\bV,\bv_j\rangle|>\varepsilon
\right]\\
&\qquad\leq
\frac{2(i-1)}{n^2}<1.
\end{aligned}
\]

Consequently, there exists at least one vector
in $\mathcal V$ satisfying all the required
inequalities, regardless of which vectors were
chosen in the previous steps.
Thus, the greedy algorithm never gets stuck.

Finally, there are $2^k$ candidate vectors,
and each candidate can be checked against all
previously chosen vectors in $O(nk)$ time.
Since there are $n$ iterations, the total
running time is
\[
O(n^2k2^k)=n^{O(\varepsilon^{-2})}.
\]
This completes the proof.
\end{proof}

In particular, the nearly orthogonal configuration
required in the proof of \Cref{thm:main} can be
obtained deterministically without increasing
the dimension.
\end{document}